\documentclass[prl,twocolumn,10pt,superscriptaddress,showpacs,aps]{revtex4-1}

\usepackage{amsmath}
\usepackage{amssymb}
\usepackage{graphicx}
\usepackage{color}

\usepackage[T1]{fontenc}
\usepackage[latin9]{inputenc}

\usepackage{amsthm}
\usepackage{bbm}
\usepackage{mathtools}

\def\tr{\operatorname{tr}}
\def\idty{{\leavevmode\rm 1\mkern -5.4mu I}} %  unit operator

\def\Ir{{\mathbb Z}}

\def\Norm #1{\left\Vert #1\right\Vert}
\def\opNorm #1{\Norm{#1}}%_{\,{op}}} % Chris's request
\def\mod{{\mathop{\rm mod}\nolimits}}

\def\ket #1{\vert#1\rangle}

\def\tr{\mathop{\rm tr}\nolimits}
\def\abs#1{\vert#1\vert}

\def\Order{{\mathcal O}}

\def\EF{\Phi} %electric field, originally E
\def\nr{n} %numerator in cf expansion
\def\dr{m} %denominator in cf expansion
\def\We{W_{\EF}}  %electric walk
\def\kk{i}   %cf expansion index - n_n looks bad...
\begin{document}

\title{Revivals in Quantum Walks with quasi-periodically time-dependent coin}

\author{C. Cedzich}
\affiliation{Institut f\"ur Theoretische Physik, Leibniz Universit\"at Hannover, Appelstr. 2, 30167 Hannover, Germany}
\author{R. F. Werner}
\affiliation{Institut f\"ur Theoretische Physik, Leibniz Universit\"at Hannover, Appelstr. 2, 30167 Hannover, Germany}

\begin{abstract}
We provide an  explanation of recent experimental results of Xue et al.\ \cite{susi}, where full revivals in a time-dependent quantum walk model with a periodically changing coin are found.  Using methods originally developed for ``electric'' walks with a space-dependent, rather than a time-dependent coin, we provide a full explanation of the observations of Xue et al. We extend the analysis from periodic time-dependence to quasi-periodic behaviour with periods incommensurate to the step size. Spectral analysis, one of the principal tools for the study of electric walks, fails for time-dependent systems, but we find qualitative propagation behaviour of the time-dependent system in close analogy to the electric case.
\end{abstract}


\maketitle

\section{Introduction}

Quantum walks are fundamental dynamical systems, involving a walking particle with internal degrees of freedom moving in discrete time steps on a lattice \cite{Ambainis2001,Grimmett,SpaceTimeCoinFlux,GenMeasuringDevice}. They have become an important test bed for many complex quantum phenomena, being well accessible both to experimental and theoretical investigation. In particular, they have recently attracted much attention as a computational resource \cite{AmbainisKR05,Magniez:2011ke,Anonymous:rSnqW2vc,Lovett:2010ff,SearchOnFractalLattice,Quenching}. They exhibit a rich variety of quantum effects such as Landau-Zener tunneling \cite{landauzehner}, the Klein paradox \cite{kurzwel:2008dm} and Bloch oscillations \cite{Gensketal}. By taking into account on-site interactions between two particles performing a quantum walk, the formation of molecules has also been established \cite{molecules,moleculestheothers,moleculestheothers2}. Recently, a complete topological classification of quantum walks obeying a set of discrete symmetries has been obtained \cite{Kita,Kita2,KitaObservation,Asbo1,Asbo2,Asbo3,Asbo4,UsOnTop,LongVersion}.
Quantum walks have been experimentally realized in such diverse physical systems as neutral atoms in optical lattices \cite{Karski:2009}, trapped ions \cite{Zaehringer:2010bs,Schmitz2009}, wave guide lattices \cite{Peruzzo:2010co,PhysRevLett.108.010502} and light pulses in optical fibres \cite{Schreiber:2010cl,Schreiber:2012ix} as well as single photons in free space \cite{PhysRevLett.104.153602}.
On the other hand, one can also observe a growing interest in quantum walks in mathematical literature where they are viewed as physical realizations of the more abstract concept of CMV matrices, the unitary analog to Jacobi matrices \cite{CMVoriginal,CGMV,recurrence,QDapproach,Damanik}.

It is easy to make the coin operation of a quantum walk depend on either the location of the walker or the time (number of time step) or both. A complete analysis for the randomly time-dependent case was given in Ref.~\cite{TRcoin}, even when the coin choice is driven by an external Markov process, hence allowing for correlated coin choices. In this case, the ballistic propagation ($x(t)\sim t$) of the non-random system reverts to diffusive propagation $x(t)\sim t^{1/2}$, i.e., Gaussian spreading with a momentum-dependent diffusion constant. In sharp contrast, disorder in space (i.e., a space-dependent set of coins fixed throughout the evolution) leads to Anderson localization \cite{Localization,dynlocalain}, i.e., purely discrete spectrum with exponentially localized eigenfunctions, and no propagation. Combining both kinds of disorder \cite{TRcoin} again leads to diffusive scaling, so adding temporal disorder will slow down propagation in a non-random system but will speed up an Anderson localized one.

It is well-known that quasi-periodic potentials share the possibility of Anderson localization with disordered ones. For quantum walks this has been analyzed in detail in \cite{electric,InhomogeneousWalkFillman}. In \cite{electric} the critical parameter is the electric field $\Phi$. For rational fields $\Phi=2\pi n/m$ one observes sharp revivals after $m$ or $2m$ steps, which are exponentially sharp as a function of $m$. Hence, somewhat paradoxically, the revival is the sharper the longer it takes. On the other hand, the evolution does not become exactly periodic, and small errors accumulate over revival cycles leading ultimately to ballistic transport. For irrational fields sufficiently close to a rational, i.e., up to ${\mathcal O}(1/m^2)$ as for the continued fraction convergents, one also sees the revivals. However, depending on the infinite sequence of convergents, the long term evolution may be quite different.  It may involve further, yet sharper revivals on larger time scales, but typically (with probability one) localization-like behaviour.

Looking at quasi-periodic temporal modulations of the dynamics as in \cite{susi} is a natural question. However, judging from the experience with random choices one would hardly expect the same methods to apply. Yet this is the case, as we show in this letter. The core of the argument is again a revival statement for the rational case, based on a trace formula established in \cite{electric}. However, the reason for the appearance of continued fraction approximations is different in the two cases. For example, in the temporal case analyzed in this note the revival statement holds uniformly for all initial wave functions, whereas in \cite{electric} we had to restrict the initial support region. Another marked difference is that in the temporal case we are not repeatedly applying the same unitary operator, so there is no operator of which we could gather (discrete vs.\ continuous) spectral information. Therefore, the interplay between spectral properties and propagation properties, which is typical for autonomous (not explicitly time-dependent) evolutions, has no analogue in the temporal case.

\section{The system}

We consider, as in \cite{susi}, a translation invariant quantum walk on the 1D lattice $\Ir$ with local spin-1/2 degree of freedom. Basis vectors of the system Hilbert space are thus denoted by $\ket{x,s}$ with $x\in\Ir$ and $s=\pm1$. The standard state-dependent shift $S$ acts as $S\ket{x,s}=\ket{x+s,s}$ and $C(t)$ is a time-dependent coin acting solely on $s$, the internal degree of freedom.  The concrete model given in \cite{susi} is given by
\begin{equation}\label{eq:coin}
  C(t)=R_x(t \EF)R_y(\theta),
\end{equation}
with $R_{x,y}(\theta)$ the rotation around the $x$ and $y$ axis in spin space, respectively. The $t^{\rm th}$ time step of the walk is then given by the unitary operator $\We(t)=SC(t)$. Note that when $\Phi/(2\pi)=n/m$ is rational, we have $W(t+k*m)=W(t)$ so the evolution is periodic. Otherwise, it is quasi-periodic. In either case we use
\begin{equation}\label{Wm}
 \We^{[m,1]}=\We(m)\We(m-1)\cdots\We(1)
\end{equation}
as a short hand for the first $m$ steps of the walk. For the rest of the paper we will generalize the above coin operator \eqref{eq:coin} by allowing instead of $R_y(\theta)$ a slightly more general unitary coin such that the system under consideration becomes
\begin{equation}\label{eq:we}
  \We(t)=SR_x(t\EF)\begin{pmatrix}a&b\\-b^*&a^*  \end{pmatrix},
\end{equation}
where $\abs{a}^2+\abs{b}^2=1$. All results remain valid if instead of $R_x(t\EF)$ we would have chosen any unitary $R$ with $R^m=\idty$ for some $m\in\mathbb N$. The choice $R=R_x(\EF)$ is made to retain analogy with \cite{susi}.

The electric walk turns out to be closely related. It has no time dependence in the coin operation (so $\Phi=0$ in \eqref{eq:coin}). Instead, after each step the wave function is multiplied by the $x$-dependent phase $\exp(i \Phi x)$, where $\Phi$ plays the role of an electric field \cite{electric}.

The basic observation in \cite{susi} is that for certain fine-tuned rational values of $\EF,\theta$ and a specific initial state there are revivals (see \cite[Table I, Table II]{susi} for theoretical and experimental results, respectively). This observation will be generalized in this manuscript to almost all values of $\EF,\theta$. These revivals will no longer be exact, so that, in contrast to \cite{susi} the evolved state will not be exactly periodic. When $\EF/(2\pi)$ is rational the denominator of $\EF/(2\pi)$ sets the time for the revival, which will even be exponentially sharp in the denominator. Hence even for a moderately large denominator the time evolution will be periodic for all practical purposes. For irrational parameters one typically still gets an infinite hierarchy of revivals governed by the continued fraction expansion of $\EF/(2\pi)$. Remarkably, these qualitative features are independent of $\theta$ and the initial state. Similar behaviour is known in the case where $\EF$ is an electric field \cite{electric}, and indeed the analysis of the rational case uses a formula originally developed for that case.

\section{Revivals and a trace formula}

\begin{figure}
  %\begin{tabular}{cc}
  %\includegraphics[height=.25\textheight]{bloch.pdf}&
  \begin{center}
  \includegraphics[width=.85\columnwidth]{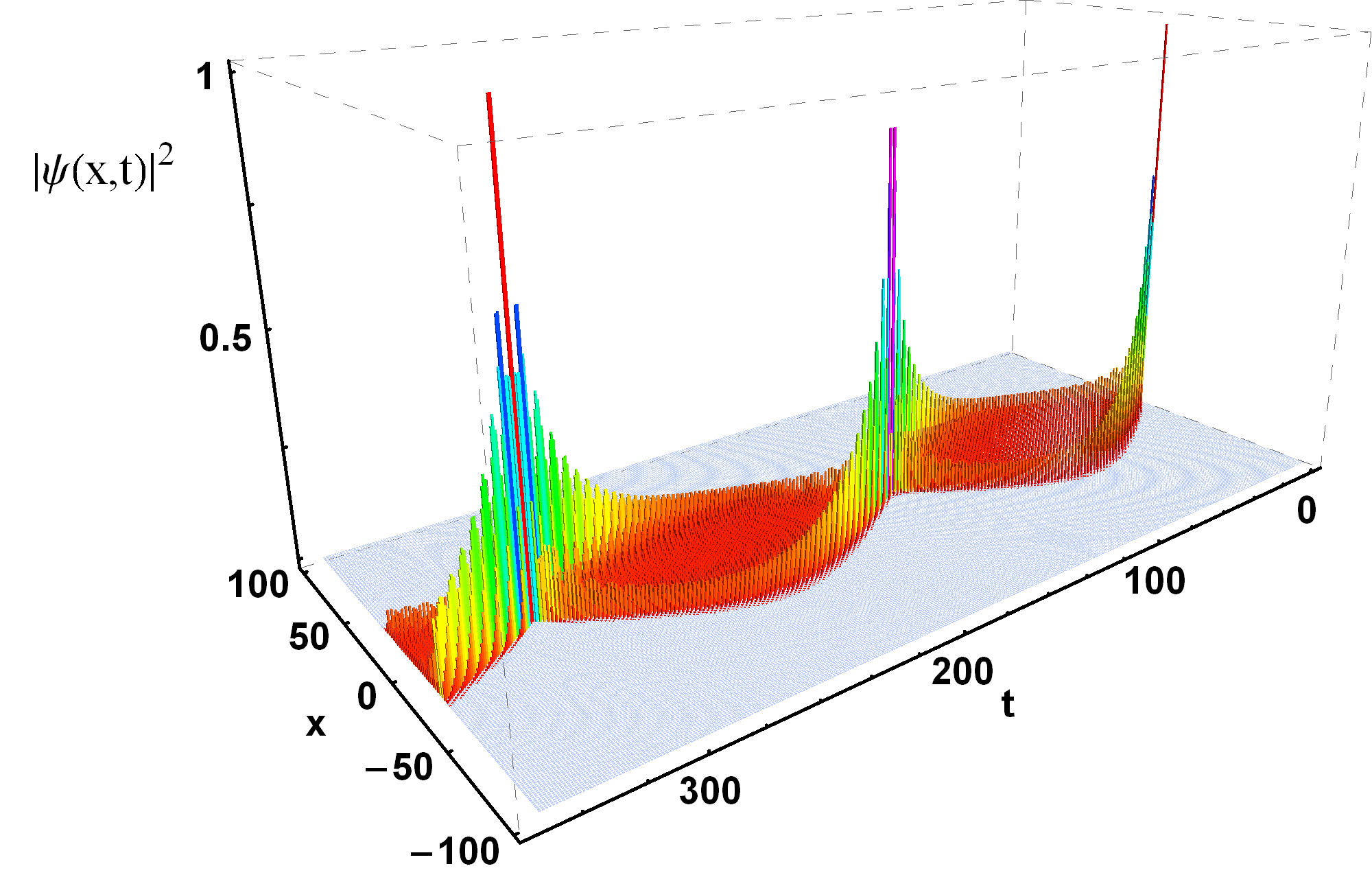}
  \end{center}
  %\end{tabular}
  \caption{%Left: Bloch sphere representation of a time evolved state as a function of quasi-momentum. The initial state was $\psi=\ket{0}$, subject to 7 steps for $\EF=2\pi/7$ and the coin corresponding to the standard Hadamard coin. \\Right:
  Position distributions for $\EF=2\pi/155$ and $\phi=\pi/4$. Note the expected revival at $t=310$ and the Bloch oscillations described in \cite{Gensketal}.}
  \label{fig:revivals}
\end{figure}

We begin the analysis with the rational case $\EF=2\pi\nr/\dr$ with $\nr$ and $\dr$ coprime. In this case $R_x^t(\EF)$ is periodic in $t$ with $R_x^m(\EF)=\idty$. Our main result is the so called \emph{revival theorem} similar to that in \cite{electric}. It states that for $\dr$ odd $\We^{[2\dr,1]}$ and for $\dr$ even $\We^{[\dr,1]}$ are norm-close to the identity with quality exponentially good in $m$:
\begin{alignat}{2}\label{reviveOdd_app}
    \opNorm{\,\We^{[2\dr,1]}+\idty\,}&= &\: \mathcal O(\abs{\widetilde\alpha}^m)\qquad&\dr\text{ odd} \\
    \opNorm{\,\We^{[\dr,1]}+(-1)^{\dr/2}\idty\,}&= &\: \mathcal O(\abs{\widetilde\alpha}^m)\qquad &\dr\text{ even}\;
    \label{reviveEven_app}
\end{alignat}
where $\widetilde\alpha$ depends solely on the coin parameters. The exponentially good quality of the estimates depends on $\abs{\widetilde\alpha}<1$. For the Hadamard walk with $a=b=1/\sqrt{2}$ the deviation from a perfect reproduction of the initial state is $2^{-m/2+1}$. The exponentially good quality of the revivals for this choice of coin are illustrated in Figure~\ref{fig:revivals} for $m=310$. In sharp contrast, for coin parameters $a=i/\sqrt{2}$ and $b=1/\sqrt{2}$ we find $\abs{\widetilde\alpha}=1$ and hence no revival predictions at all. Also, the difference in behavior for $\dr$ even and odd is understood intuitively, since the probability to find the particle at the origin is non-zero only after an even number of steps. %Hence the revival of a state localized at a point is only possible after an even number of steps.
To prove the above revival theorem note that the ``temporally regrouped'' walk $\We^{[\dr,1]}$ is independent of time. Hence we can apply the standard theory of translation invariant walks (see, e.g. \cite{Grimmett,SpaceTimeCoinFlux,TRcoin}) and consider the Fourier-transformed operator
\begin{eqnarray}\label{WEnp}
    \We^{[\dr,1]}(k)&=&S(k)R_x(\EF)CS(k)R_x^2(\EF)\dots R_x^m(\EF)C
\end{eqnarray}
in momentum space with dispersion relation
\begin{equation}\label{eq:disrel}
  2\cos\omega_\pm(k)=\tr\We^{[\dr,1]}(k).
\end{equation}

To get an explicit expression for $\omega_\pm$ we thus need to evaluate the trace $\tr\We^{[\dr,1]}(k)$ for which we adapt a result from \cite{electric},
called the \emph{trace formula}. This provides an expression for traces of the type $\tau_m(M) = \tr(MR^0MR^1\cdots MR^{m-1})$ where $R^m=\idty$ with $R$ unitary 
and $M=\left(\begin{smallmatrix}  \alpha    & \beta\\ \gamma & \delta \end{smallmatrix}\right)$ is a general $2\times2$ matrix. For $m$ odd we get:
\begin{equation}\label{trodd}
 \tau_m(M)=\widetilde\alpha^m+\widetilde\delta^m\;,
\end{equation}
and for $m$ even:
\begin{equation}\label{treven}
 \tau_m(M)=-\bigl(\widetilde \alpha^{m}+\widetilde \delta^{m}\bigr)+2(-1)^{m/2}\bigl((\widetilde \alpha\widetilde \delta)^{m/2}-\det(M)^{m/2}\bigr).
\end{equation}
Here, $\widetilde\alpha  =(BMB^*)_{11}$ and $\widetilde\delta  =(BMB^*)_{22}$ and $B$ is the unitary diagonalizing $R$.

We apply the trace formula to the walk \eqref{WEnp} with rational $\EF/(2\pi)=\tfrac\nr\dr$ such that $R_x^\dr(\EF)=\idty$. In \eqref{trodd} and \eqref{treven} we let
$M\mapsto CS(k)$ which corresponds to replacing $\alpha\mapsto ae^{ik},\beta\mapsto be^{-ik},\gamma\mapsto -b^*e^{ik},\delta\mapsto a^*e^{-ik}$ and $\det M\mapsto1$. Note that $R\propto R_x=\exp(i\sigma_x)$ in the trace formula requires $B=\tfrac1{\sqrt{2}}\left(\begin{smallmatrix}1&1\\1&-1\end{smallmatrix}\right)$ which yields $2\widetilde\alpha=\alpha+\beta+\gamma+\delta$ and $2\widetilde\delta=\alpha-\beta-\gamma+\delta$. Writing $a=\abs ae^{ik_a}$ and $b=\abs be^{ik_b}$ in polar form then results in
\begin{equation*}
  \widetilde\alpha  =\widetilde\delta^*=\abs{a}\cos(k_a+k) + i\abs{b}\sin(k_b-k)
%  \widetilde d  &=\abs{a}\cos(k_a+k)+i\abs{b}\sin(k_b+\red{-}k).
\end{equation*}
such that by writing $\widetilde\alpha=\abs{\widetilde\alpha}\exp(i\theta_{\widetilde\alpha})$
%with $\abs{\widetilde a}^2=\abs{a}^2\cos^2(k_a+k)+\abs{b}^2\sin^2(k_b+k)$ and $\tan(\theta_{\widetilde a})=\frac{\abs{b}\sin(k_b+k)}{\abs{a}\cos(k_a+k)}$
the dispersion relation \eqref{eq:disrel} reads
\begin{align}\label{dispersEpaps}
  &\cos\omega_\pm(k)=\\
  &=\begin{cases}\abs{\widetilde\alpha}^m\cos(m\theta_{\widetilde\alpha})& m\mbox{\ odd}\\
                               -\abs{\widetilde\alpha}^m\cos(m\theta_{\widetilde\alpha})+2(-1)^{\frac m2+1}\left(1-\abs{\widetilde\alpha}^m\right) &  m\mbox{ even}.
%  &=\begin{cases}\left(\abs{a}\cos(k_a+k) - i\abs{b}\sin(k_b+k)\right)^m+\left(\abs{a}\cos(k_a+k)+i\abs{b}\sin(k_b+k))\right)^m& m\mbox{\ odd}\\
%                               -\left(\abs{a}\cos(k_a+k) - i\abs{b}\sin(k_b+k)\right)^m-\left(\abs{a}\cos(k_a+k) + i\abs{b}\sin(k_b+k)\right)^m-2\left(-\left(\abs{a}^2\cos^2(k_a+k)+\abs{b}^2\sin^2(k_b+k)\right)^{\tfrac m2}-1\right) &  m\mbox{ even}
%                              (-1)^{{m}/2+1}(1-2a^{{m}/2}\cos({m}k/2))  &  m\mbox{ even}
                              \end{cases} \nonumber
\end{align}
Using this expression in the proof of the revival theorem in \cite{electric} yields Eqs. \eqref{reviveOdd_app} and \eqref{reviveEven_app}.

\section{The irrational case}

The next step is to consider irrational values for $\EF/(2\pi)$. Here, the spectral picture breaks down due to the time-dependence of the operator $\We(t)$ as there is no concatenation of $\We(t)$ which is periodic in $t$. However, one may still classify such systems by their long-time propagation behaviour.
We distinguish between two different regimes of irrationality depending on the approximability of $\EF/(2\pi)$ by continued fractions. Denoting by $c_\kk$ the continued fraction coefficients of an irrational number $x$ and by $\nr_\kk/\dr_\kk$ its continuants we have $\abs{x-\nr_\kk/\dr_\kk}<c_{\kk+1}^{-1}\dr_\kk^{-2}$ \cite{Hardy} and it is this quadratic quality of approximation in $\dr_\kk$ which is crucial for our result. We then distinguish two regimes by how rapidly we can approximate $x$ depending on the sequence of continued fraction coefficients $c_i$. The two regimes are irrationals the sequence of $c_i$ of which is bounded or unbounded, respectively. Independent of this distinction we may estimate the norm difference of two time-dependent walks \eqref{eq:we} with fields $\EF,\EF'$ by $\opNorm{\We(t)-W_{\EF'}(t)}\leq t\abs{\EF-\EF'}$ such that, irrespective of the initial state, after $t$ steps we find
%For electric quantum walks, these approximations were used to prove that any \emph{finitely localized} intital state revives at times equal to the denomiators of the continuants. There, the localized initial state was necessary due to the linear spacial dependence of the electric field which allowed for initial states which after application of walks with infinitesimally close fields would differ by an arbitrary large amount. In the present case we do not observe such a spacial dependence. Instead we may even estimate the norm difference of two time-dependent walks \eqref{eq:we} with fields $\EF,\EF'$ by $\Norm{\We(t)-W_{\EF'}(t)}\leq t\abs{\EF-\EF'}$ such that, irrespective of the initial state, for its $t^{\text{th}}$ concatenation we find
\begin{equation*}
  \opNorm{\We^{[t,1]}-W_{\EF'}^{[t,1]}}\leq \frac t2(t+1)\abs{\EF-\EF'}.
\end{equation*}
Taking $\EF'/(2\pi)=\nr_i/\dr_i$ to be a continuant of $\EF$ we find, due to the quadratic quality of the approximation of $\EF$ in $\dr_i$,
\begin{equation}\begin{aligned}\label{reviveirr}
    \opNorm{\,\We^{[2{\dr}_{\kk},1]}+\idty\,}&\leq \frac{4\pi}{c_{\kk+1}}+\Order\left(\tfrac1{{\dr}_{\kk}}\right) \quad\dr_{\kk} \text{ odd}   \\
    \opNorm{\,\We^{[\dr_{\kk},1]}+(-1)^{\dr_{\kk}/2}\idty\,}&\leq \frac{\pi}{c_{\kk+1}}+\Order\left(\tfrac1{\dr_{\kk}}\right)  \quad\dr_{\kk}\text{ even}.
\end{aligned}\end{equation}

Thus, for irrational numbers the sequence of continued fraction of which coefficients diverges, i.e., $c_i\rightarrow\infty$, we get an infinite sequence of sharper and sharper revivals followed by farther and farther excursions. These revivals are, in contrast to \cite{electric}, independent of the initial state. Depending on the parity of the denominator of the continuants of $\EF/(2\pi)$, these revivals occur at times $\dr_i$ and $2\dr_i$ which grow at least exponentially \cite{Hardy}.

\begin{figure*}
  \begin{tabular}{lr}
  \includegraphics[height=.22\textheight]{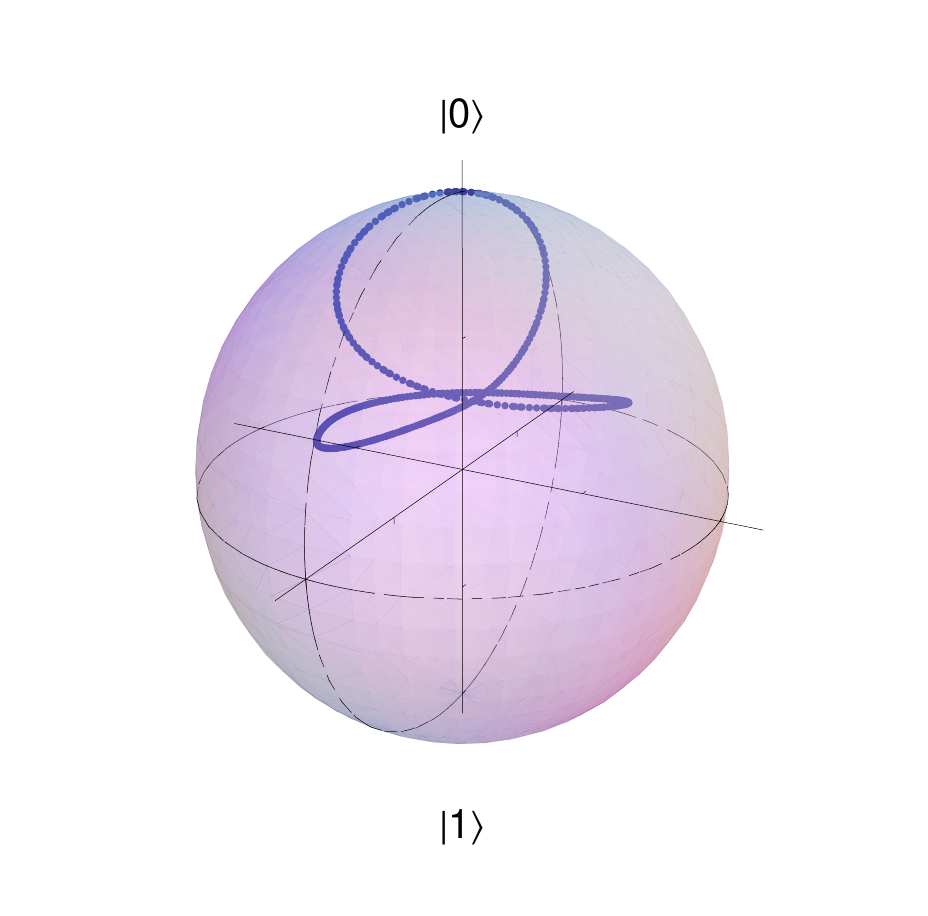}&\includegraphics[height=.22\textheight]{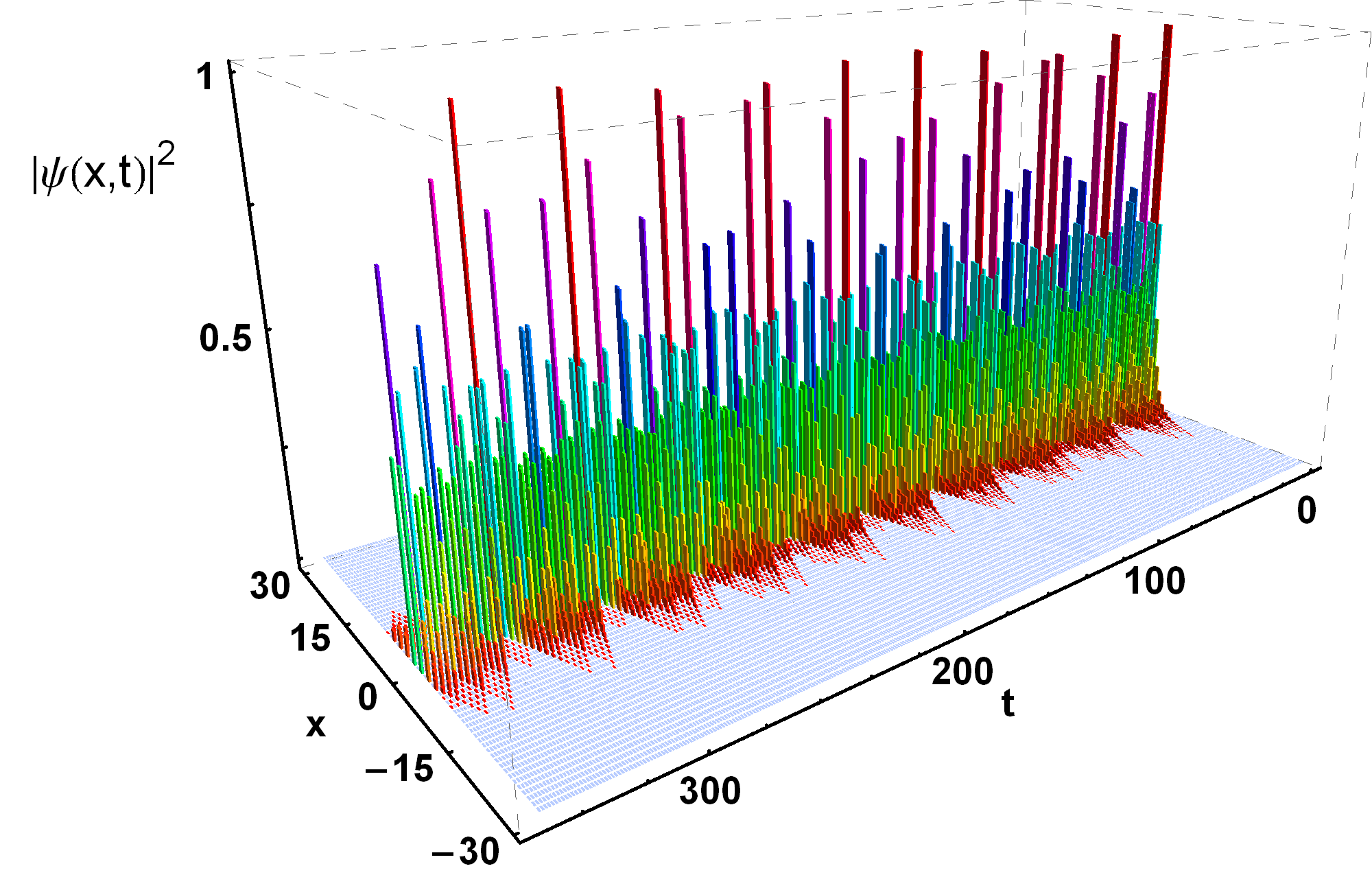}
  \end{tabular}
  \caption{\label{fig:irrat}
  Left: Bloch sphere representation of the $x=0$ component of the wave function $\psi(x,t)$ as a function of $t$ with $t$ up to 1000 with $\EF=2\pi\varphi$ and the coin corresponding to the standard Hadamard coin. Despite the appearance of a smooth curve, successive values of $t$ are not close to each other, but rather the curve is sampled in an incommensurate way. \\
  Right: Position distributions for the Golden Ratio $\EF=2\pi\varphi$ and $\phi=\pi/4$. Erratically appearing revivals not showing any periodicity are clearly visible. Also, the initial state does not spread more than approximately 15 sites which suggests that systems with quasi-periodic time dependence do not exhibit any transport.}
\end{figure*}

For numbers the sequence of continued fraction coefficients of which is bounded, however, \eqref{reviveirr} does not predict any revivals. The best known and worst approximable irrational is the Golden Ratio $\EF/(2\pi)=\varphi=(\sqrt{5}-1)/2$, which has constant continued fraction coefficients $c_i=1$. Numerical simulations suggest that such systems do not show transport at all - a conjecture similar to that in \cite{electric} the analytical proof of which is work in progress. Figure~\ref{fig:irrat} shows the trajectory of the initial vector $\psi=\ket0\otimes\ket0$ on the Bloch sphere and the position distribution for $\EF/(2\pi)=\varphi$. The trajectory of the Bloch vector at $x=0$ follows a closed curve which, after delving into the interior of the Bloch sphere approaches the initial vector $\psi$ arbitrarily close at times independent of any continued fraction of $\varphi$ due to the quasi-periodic dependence of the walk on $\EF$. These erratic revivals of arbitrarily good quality (Fig.~\ref{fig:irrat}, left) and such behaviour in time-independent systems would be a signature of pure point spectrum with exponentially decaying eigenfunctions, in the literature referred to as \emph{Anderson localization}. However, as noted above, such a spectral treatment is meaningless in the time-dependent case. Additionally, irrationals with bounded continued fraction coefficients are of measure zero \cite{Hardy}, such that even in the irrational case we proved the occurrence of revivals at times equal to the denominators of the continuants for almost all $\EF$.

\section{Impurities in the choice of $\EF$}

As experimental implementations never are perfect but always contain impurities, let us briefly comment on the validity of the results above in the presence of (linear) noise in $\EF$. Exact rationality of this parameter seems necessary for the appearance of the revivals as this exactness is indeed for stating the revival theorem.
%Experimental noise is best modeled by drawing $\EF$ independently in each timestep according to the Gaussian probability distribution on $I(\EF_0)=[\EF_0-\pi,\EF_0+\pi]$ around some $\EF_0$. The measure according to which we chose is given by
%\begin{equation}
%  \nu(X)=\frac1N\int_Xe^{-\frac{(\EF-\EF_0)^2}{\sigma^2}}d\EF
%\end{equation}
%where $X$ is any interval in $I(\EF_0)$ and $N$ is a normalization constant such that $\nu(I(\EF_0))=1$. Then, for estimating the revivals we have to consider
%\begin{equation}
%  \mathbb E_\nu(\Norm{W_\EF-W_\nu})
%\end{equation}
%where we denoted by

\begin{figure}[t]
  \includegraphics[width=\columnwidth]{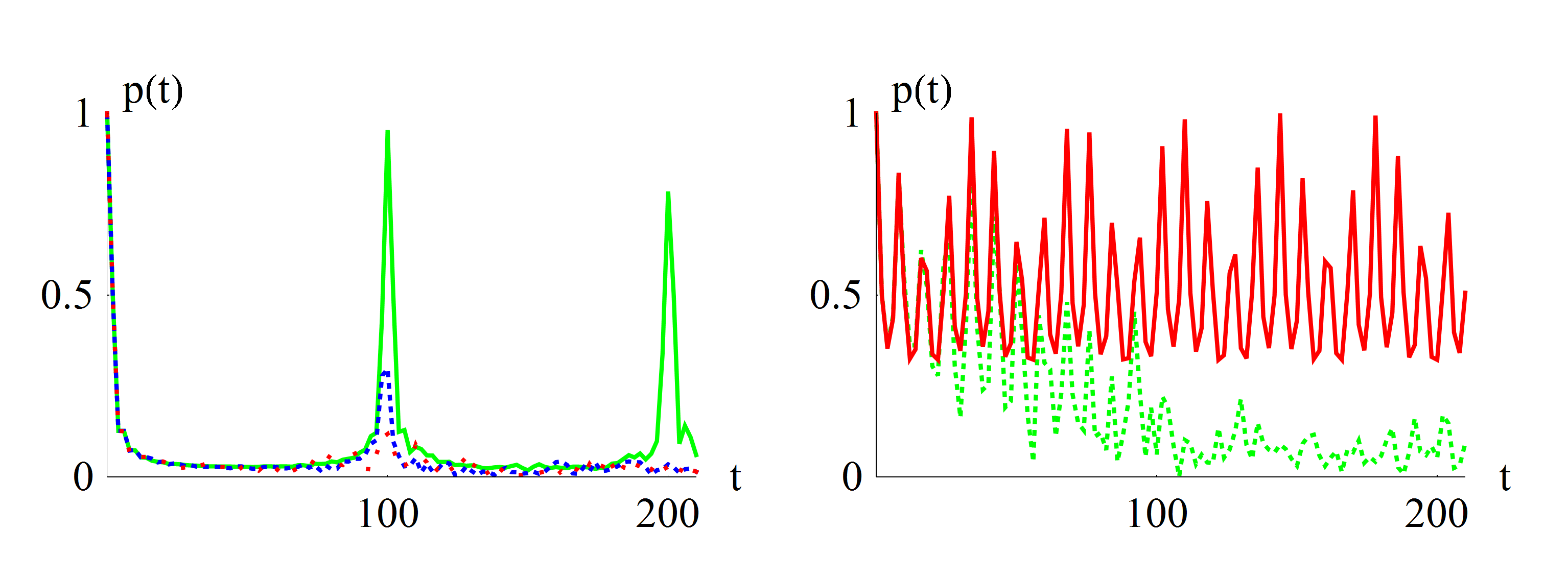}
  \caption{Return probabilities $p(t)$ for $\EF_\epsilon$ chosen randomly in each time step. Left: $\EF=2\pi/100$ where the green (solid), blue (dashed) and red (dotted) lines correspond to noise $\epsilon=10^{-4},5\times10^{-4}$ and $\epsilon=10^{-3}$, respectively. The revivals at $t=100$ in the first two cases as well as the absence of any revival for $\epsilon=10^{-3}>\Order(100^{-2})$ are predicted by \eqref{eq:noiseestimates}. Right: $\EF=2\pi(\sqrt{5}-1)/2$. Whereas $p(t)$ for $\epsilon=0$ (red, solid) is bounded from below, it converges to zero for $\epsilon=10^{-3}$ (green, dashed), thereby implying transport.}
  \label{fig:probsgrid}
\end{figure}

However, as already may be inferred from the estimate for irrational values of $\EF$, there is some stability of the revivals against noise in $\EF$. Let us model fluctuations by
\begin{equation*}
  \EF_\epsilon(x_t)=\EF+\epsilon x_t,\qquad x_t\in[0,1],
\end{equation*}
where $x_t\in[-1,1]$ is chosen randomly in each time step. Using the approximation above with $\EF'=\EF_\epsilon(x_t)$ we find
\begin{equation*}
  \opNorm{\We^{[t,1]}-W_{\EF'}^{[t,1]}}\leq \tfrac t2(t+1)\epsilon.
\end{equation*}
Thus in analogy with \eqref{reviveirr} for $\EF/(2\pi)=\nr/\dr$
\begin{equation}\begin{aligned}\label{eq:noiseestimates}
    \opNorm{\,W_{\EF'}^{[2\dr,1]}+\idty\,}&\leq\dr(2\dr+1)\epsilon+\Order(\widetilde\alpha^\dr) \\
    \opNorm{\,W_{\EF'}^{[\dr,1]}+(-1)^{\dr/2}\idty\,}&\leq\tfrac\dr2(\dr+1)\epsilon+\Order(\widetilde\alpha^\dr)
\end{aligned}\end{equation}
for $\dr$ odd and even, respectively. Hence if random fluctuations can be controlled on the order of $\epsilon=\Order(\dr^{-2})$ signatures of revivals are found, see Figure~\ref{fig:probsgrid}.

Quite striking is the observation that for irrational values of $\EF$ for which no transport is observed in clean systems, such as the Golden Ratio, the presence of noise makes the walk propagate. This "noise-induced transport" is a reminiscent of the fact that the set of values showing no transport has measure zero such that independent of $\epsilon>0$ with probability one in each step the random variable $\EF_\epsilon$ induces transport (Figure~\ref{fig:probsgrid}, right).

%\red{Of course, in laboratories fluctuations usually are not uniformly distributed on some interval but rather according to the normal distribution $\mathcal N(\EF,\sigma^2)$ with variance $\sigma^2$. The statements above, however, remain the same, i.e. $\sigma^2$ has to be controlled on the order of $\dr^{-2}$ such that this case}

\section{Gauge-equivalence between walks quasiperiodic in space and time}

The similarity of the results in the body of the paper and those for electric walks \cite{electric} strongly suggest a relation between the two models. In continuous time on the lattice systems with linear potential and systems with a uniform time-dependent vector potential are gauge-equivalent.Led by this example we here examine the possibility of transforming the spatial dependence of electric walks to a temporal one by a local gauge transformation $G_t=\bigoplus_xG_{x,t}$. The electric walk model is defined by
\begin{equation*}\label{eq:welectric}
  W^E_\EF=e^{i\EF\hat x}CS=:\bigoplus_x C_xS,
\end{equation*}
where $\hat x$ denotes the position operator. To establish a gauge equivalence between $W^E_\EF$ and an explicitly time-dependent walk $W(t)$ (like the one in \eqref{eq:we}) we have to find $G_{x,t}$ such that
\begin{equation*}
  W(t):=G_{t}W^E_\EF G_{t-1}^*
\end{equation*}
is uniform in space, i.e. translation invariant, but explicitly time-dependent. By unitarity of $G_t$ we find
\begin{equation*}
  W^{[t,1]}=W(t)\dots W(1)=G_t(W^E_\EF)^tG_0.
\end{equation*}
Demanding locality of the $G_{x,t}$ and a shift-coin decomposition of $W(t)$ forces $G_{t}$ and $S$ to commute, which is guaranteed by choosing the ansatz $G_{x,t}=e^{-i\EF tx}\idty$. Then we find the time-dependent and translation invariant walk
\begin{equation}\label{eq:gauged}
  W(t)=Ce^{-i\EF(t-1)\sigma_z}S.
\end{equation}
Comparing this time-dependent walk with \eqref{eq:we} we find that though the models are not exactly gauge-equivalent, the operators implementing time-dependence $e^{-i\EF(t-1)\sigma_z}$ in \eqref{eq:gauged} and $R_x(t\EF)$ in \eqref{eq:we} are unitarily equivalent, since the occurrence of revivals qualitatively does not depend on the explicit form of the time-dependent operator $R$ but only on the condition $R^\dr=\idty$ for some $\dr\in\mathbb N$. Quantitatively, however, the revival structure does depend on $R$ via $\widetilde\alpha$. The revival predictions for electric walks hence agree with those for \eqref{eq:gauged} but not for those for \eqref{eq:we}. Note that these revivals for \eqref{eq:gauged} in the irrational case become independent of the support of initial states in sharp contrast to \cite{electric}. This is a reminiscent of the spatial dependence of the $G_{x,t}$.

\section{Conclusion and Outlook}

In spite of the fact that the modification of quantum walks by randomness in time and by randomness in space leads to qualitatively very different phenomena we have established a case where quasi-periodic modifications in space and time, respectively, lead to very similar behaviour, especially with regard to revivals. This holds for both the commensurate and the incommensurate case and revival signatures are shown to be stable with respect to noise in the quasi-periodic parameter. The models studied here are based on coined quantum walks with qubit coins, a fact that is used in an essential way. There appears to be no general mapping allowing such conclusions in more general cases. Indeed some similarities observed numerically, such as the appearance of a ``smooth'' trajectory in Fig.~\ref{fig:irrat} (left), which is well understood in the space-quasi-periodic case await an analytic explanation. It would also be very interesting to establish connections for systems with higher dimensional lattices and coin spaces.

\section*{Acknowledgements}

The authors thank Referee A for his suggestion to consider a gauge equivalence between the model in this paper and the electric walk model. We acknowledge support from the ERC grant DQSIM and the European project SIQS.

\appendix*

\section{APPENDIX: Detailed investigation of \cite[Table I]{susi}}

We discuss the findings of \cite[Table I]{susi} by means of the more general theory laid out in the body of the paper. In \cite{susi} the coin is given by $C=R_y(\theta)$ and results are given for $\theta=0$ and $\theta=\pi/2$ \cite[Table I]{susi}. These choices lead to $C=\idty$ and $C=-i\sigma_y$, respectively, for which the revival theorem predicts perfect revivals as we show below. Let us first examine the case $C=\idty$. Then
\begin{equation*}
  \abs{\widetilde\alpha}^2=\cos^2(k),\qquad \cos(m\theta_{\widetilde\alpha})=1
\end{equation*}
such that the revival theorem \eqref{reviveOdd_app} and \eqref{reviveEven_app} results in
\begin{align*}
  \opNorm{\We^{2m}+\idty}&=2,\quad &m\:\text{odd}, \\
  \opNorm{\We^m+(-1)^{m/2}\idty}&=0,\quad &m\:\text{even}.
\end{align*}
Hence, contrary to the electric case where $C=\idty$ shows no revivals at all, for $m$ even periodically occurring perfect revivals are predicted. For $m$ odd no revivals are predicted which agrees with \cite[Table I]{susi} where only even values for $m$ are considered. Choosing $C=i\sigma_y$ yields
\begin{equation*}
  \abs{\widetilde\alpha}^2=\sin^2(k),\quad \cos(m\theta_{\widetilde\alpha})=\begin{cases}1  &   m\mod 4=0   \\ 0    &   m\:\text{odd}   \\ -1 & m \mod4=2\end{cases}
\end{equation*}
such that in either case
\begin{align*}
  \opNorm{\We^{2m}+\idty}&=0,\quad &m\:\text{odd}, \\
  \opNorm{\We^m+(-1)^{m/2}\idty}&=0,\quad &m\:\text{even}
\end{align*}
giving revival predictions independent of the parity of $m$. This agrees with\cite[Table I]{susi} where for $\theta=\pi/2$ also odd values for $m$ are admissible.

\bibliography{peng}

\end{document}